\documentclass[showpacs,eqsecnum,aps,twocolumn,floatfix]{revtex4}
\usepackage{graphicx,subfigure,amsmath,amssymb}
\usepackage{epsfig}
\usepackage{multirow}
\usepackage{rotating}

\begin{document}

\title{Shape phase mixing in critical point nuclei}

\author{R. Budaca$^{1}$ and A. I. Budaca$^{1}$}

\affiliation{$^{1)}$"Horia Hulubei" National Institute for Physics and Nuclear Engineering, Str. Reactorului 30, RO-077125, POB-MG6, M\v{a}gurele-Bucharest, Romania}

\begin{abstract}
Spectral properties of nuclei near the critical point of the quantum phase transition between spherical and axially symmetric shapes are studied in a hybrid collective model which combines the $\gamma$-stable and $\gamma$-rigid collective conditions through a rigidity parameter. The model in the lower and upper limits of the rigidity parameter recovers the $X(5)$ and $X(3)$ solutions respectively, while in the equally mixed case it corresponds to the $X(4)$ critical point symmetry. Numerical applications of the model on nuclei from regions known for critical behavior reveal a sizable shape phase mixing and its evolution with neutron or proton numbers. The model also enables a better description of energy spectra and electromagnetic transitions for these nuclei.
\end{abstract}

\maketitle

\section{Introduction}

The single-particle degrees of freedom constitute the natural basis for any nuclear theory. However, bulk properties which are more important in medium- and heavy-mass nuclei are traditionally described by means of collective models. In particular, the Bohr-Mottelson model \cite{Bo,Bohr} (BMM) of nuclear surface oscillations provides an intuitive phenomenology as well as a geometric classification of the collective motion. Alternatively, a group theoretical description of nuclear collective properties is offered by the Interacting Boson Model (IBM) \cite{IBM}, which is basically a pair coupling model, with coherent monopole and quadrupole pairs of fermions approximated as bosons. The most general IBM Hamiltonian can be expressed in terms of the geometrical shape variables of BMM by means of an intrinsic coherent state \cite{Diep0,Diep1}. The resulted differential equation is, however, far more involved, bearing only a marginal equivalence to the Bohr Hamiltonian \cite{Gino1,Gino2,Kirson}. This is not surprising given the conceptual distinction between the two models. Indeed, whereas BMM is purely geometrical, IBM is actually a large truncation of the shell model. However, in the earlier attempts to relate the two approaches \cite{Arima0,Diep0,Diep1}, it was found that the limiting dynamical symmetries $U(5)$ \cite{Arima1}, $SU(3)$ \cite{Arima2}, and $O(6)$ \cite{Arima3}, identified as subgroup chains of the $SU(6)$ symmetry of IBM, have analogues in BMM represented by its solvable instances corresponding to the shape phases describing a spherical vibrator \cite{Bohr}, an axially symmetric rotor \cite{Bo}, and a $\gamma$-soft rotor \cite{Wilets} respectively. The search for explicit mappings between the two models in their solvable limits revealed that a large range of BMM results can be reproduced in various contraction limits of the IBM \cite{Thiam1,Thiam2,Thiam3}. The fact that, besides the microscopical upbringing, IBM can be understood also as a compactification of BMM serves as a bridge between the single-particle degrees of freedom and the purely geometrical collective variables defining the shape of the nuclear surface.

The presence of a symmetry is directly related to the exact solvability of the associated Hamiltonian, whose solutions can be indexed by many-body quantum numbers. This is a simpler explanation for the equivalence of the IBM dynamical symmetries with BMM solvable limits. Therefore, a lot of effort was directed to find other exactly solvable cases of the BMM \cite{Fort,Cejnar,BugFort} and their corresponding symmetries. As a result, it was found that variations of the BMM with a square well potential have analytical solutions adequate for a similar algebraic description of the critical points of the transitions between the aforementioned dynamical symmetries. Indeed, the solutions $E(5)$ \cite{Iachello0} and $X(5)$ \cite{Iachello1}, associated with the critical points of the transition lines $U(5)\rightarrow O(6)$ and $U(5)\rightarrow SU(3)$, are closely related to the five-dimensional Euclidean symmetry. More precisely, it is exactly realized in the former and only partially in the latter \cite{Bon01,Bon02,X4}. The elusive group structure of the $X(5)$ critical point at first glance might be ascribed to the adopted approximations. However, the same algebraic properties are found in its $\gamma$-rigid counterpart, the $X(3)$ model \cite{BonX3}, which is exactly separable and solvable but acts in a reduced three-dimensional shape phase space. Thus, regardless of the description associated with the $\gamma$ shape variable, the relation of the critical point solutions for the transition between spherical and axially symmetric shapes with the Euclidean group is invariable. This aspect together with the phenomenological compatibility between the $\gamma$-stable and $\gamma$-rigid conditions inspired a relaxation of the $X(5)$ critical point solution in terms of a $\gamma$ rigidity parameter. Basically, the measure of the $\gamma$ rigidity combines the quantum treatments of the collective excitations corresponding to the limiting shape phase spaces of $X(3)$ and $X(5)$ solutions. The intermediary situation obviously involves a mixed shape phase space, {\it i.e.}, something between three and five dimensions. In this paper we will show that some of the known critical point axially symmetric nuclei prefer this arrangement. Moreover, the degree of the shape phase mixing have a rather smooth evolution in well defined sequences of nuclei. A similar program was used to define the $X(4)$ critical point solution \cite{X4} as well as to combine some exactly separable variations of the $X(5)$ model to their $\gamma$-rigid limits \cite{Noi1,Noi2}.

\renewcommand{\theequation}{2.\arabic{equation}}
\setcounter{equation}{0}
\section{Shape phase space mixing}

The general Bohr model \cite{Bohr} for quadrupole shapes has in total five variables: two associated with the nuclear shape oscillations and three Euler angles describing the rotational motion. Restricting the $\gamma$ shape variable to certain values, we can obtain more simple models \cite{BonX3,BonZ4} due to smaller number of degrees of freedom. Indeed, the fixed $\gamma$ variable becomes a simple parameter, and the quantum Hamiltonian associated with such a case will have a different structure according to the Pauli quantization prescription \cite{Pauli}. An interesting situation arises in more restrained conditions of a prolate $\gamma$-rigid system ($\gamma=0$) \cite{BonX3}. Due to the symmetry properties, its rotational motion can be described by only two Euler angles and therefore the whole system will have just three variables instead of five as in the usual Bohr model.

The small-angle approximations made on the $\gamma$ shape variable in $\gamma$-stable models is quite similar to the $\gamma$-rigid conditions. This correspondence led to the idea of a hybrid model based on the interplay between $\gamma$-stable and $\gamma$-rigid collective excitations \cite{Noi1,Noi2,X4}. It was achieved by introducing a control parameter $0\leq\chi<1$ called $\gamma$ rigidity, which mediates a coupling between the two types of collective excitations:
\begin{equation}
H=\chi \hat{T}_{r}+(1-\chi)\hat{T}_{s}+V(\beta,\gamma),
\label{Ht}
\end{equation}
where $V(\beta,\gamma)$ is the potential energy. The usual five-dimensional kinetic operator of a $\gamma$-soft Bohr Hamiltonian reads
\begin{eqnarray}
\hat{T}_{s}&=&-\frac{\hbar^{2}}{2B}\left[\frac{1}{\beta^{4}}\frac{\partial}{\partial{\beta}}\beta^{4}\frac{\partial}{\partial{\beta}}+\frac{1}{\beta^{2}\sin{3\gamma}}\frac{\partial}{\partial\gamma}\sin{3\gamma}\frac{\partial}{\partial\gamma}\right.\nonumber\\
&&\left.-\frac{1}{4\beta^{2}}\sum_{k=1}^{3}\frac{Q_{k}^{2}}{\sin^{2}{\left(\gamma-\frac{2}{3}\pi k\right)}}\right],
\label{Ts}
\end{eqnarray}
where $Q_{k}(k=1,2,3)$ denote the three projections of the angular momentum on the principal axes of the intrinsic frame of reference. Here, $\gamma$ softness is related to the property of the system to have non-axial fluctuations around an equilibrium geometry. Further, depending on the potential energy, we can have $\gamma$-stable or $\gamma$-unstable conditions. In the first case the potential has a single localized minimum in the $\gamma$ shape variable, while in the latter it does not depend on $\gamma$ at all. This terminology is unfortunately often misused, but very clear definitions can be found in Ref.\cite{Fort}. In contradistinction, the prolate $\gamma$-rigid kinetic energy operator \cite{BonX3} defined in a three-dimensional shape phase space,
\begin{equation}
\hat{T}_{r}=-\frac{\hbar^{2}}{2B}\left[\frac{1}{\beta^{2}}\frac{\partial}{\partial{\beta}}\beta^{2}\frac{\partial}{\partial{\beta}}-\frac{Q_{1}^{2}+Q_{2}^{2}}{3\beta^{2}}\right],
\label{Tr}
\end{equation}
is associated with a potential energy with an extremely sharp $\gamma$ minimum which practically does not allow fluctuations. The lack of the third component of angular momentum in the above equation is due to the quantum mechanical restriction that the rotation cannot take place around the symmetry axis.

For the purpose of this study, $T_{r}$ and $T_{s}$ will be associated with the $X(3)$ and $X(5)$ models respectively, which share the same infinite square well shape of the separated $\beta$ potential. The differences in the quantum description of the two situations arising from different shape phase space dimensions are resolved by a suitable weighting of the shape phase metric associated with the full Hamiltonian (\ref{Ht}). The origin of this deformed shape phase space lies in the general definition of the kinetic energy of the collective Hamiltonian as a Laplacian operator in curvilinear coordinates \cite{Rohozinski}:
\begin{equation}
\hat{T}=-\frac{\hbar^{2}}{2}\nabla^{2}=-\frac{\hbar^{2}}{2}\sum_{lm}\frac{1}{J}\frac{\partial}{\partial{x^{l}}}J\bar{G}^{lm}\frac{\partial}{\partial{x^{m}}}.
\label{Pauli}
\end{equation}
$J=\sqrt{det(g)}$ is the Jacobian of the transformation from the quadrupole coordinates $\{q_{k}\}$ to the curvilinear ones $\{x^{l}\}_{l=1,5}=\{\beta,\gamma,\theta_{1},\theta_{2},\theta_{3}\}$ defined by the metric tensor:
\begin{equation}
g_{lm}=\sum_{k}\frac{\partial{q_{k}}}{\partial{x^{l}}}\frac{\partial{q_{k}}}{\partial{x^{m}}},
\end{equation}
while $G_{lm}$ is a symmetric positive-definite bitensor matrix. In the general five-dimensional Bohr model, this bitensor is just the transformation tensor $g_{lm}$ up to a common mass parameter and the kinetic operator (\ref{Pauli}) acquires the well known form of the Laplace-Beltrami operator \cite{Hicks}.  This is no longer valid if we want to introduce the rigidity dependence. However, it can be easily shown that the $\chi$ dependent weighting factor arises naturally in the definition of the $\beta$ wave function if we consider the following mass tensor components in the general collective Hamiltonian:
\begin{eqnarray}
G_{lm}&=&0,\,l\neq m,\,\,G_{\beta\beta}=B,\,\,G_{\gamma\gamma}=\frac{B}{1-\chi},\\
G_{kk}&=&\frac{4B\beta^{2}}{1-\chi\delta_{k,3}}\sin^{2}{\gamma_{k}},\,\,\gamma_{k}=\gamma-\frac{2k\pi}{3},\,\,k=1,2,3.\nonumber
\end{eqnarray}
In this way we will have, in the axial rigid limit, infinite inertial parameters for the conjugate momentum of the $\gamma$ shape variable and the angular velocity $\omega_{3}=\dot{\theta}_{3}$ around the third intrinsic axis \cite{Roho}.

\renewcommand{\theequation}{3.\arabic{equation}}
\setcounter{equation}{0}
\section{Application to $X(D)$ critical points}

As the aim of the paper is to study critical point nuclei, we will treat the Schr\"{o}dinger equation associated with (\ref{Ht}) as in case of the well known $X(5)$ model \cite{Iachello1}, where an approximate separation of $\beta$ and $\gamma$ angular variables is achieved through a series of approximations conditioned by the following separated form for the total reduced potential:
\begin{equation}
u(\beta,\gamma)=\frac{2B}{\hbar^{2}}V(\beta,\gamma)=u(\beta)+(1-\chi)v(\gamma).
\label{v}
\end{equation}
In case of a very sharp $\gamma$ potential centered around $\gamma=0$, the rotational term from (\ref{Ts}) can be very well approximated by
\begin{equation}
\sum_{k=1}^{3}\frac{Q_{k}^{2}}{\sin^{2}{\left(\gamma-\frac{2}{3}\pi k\right)}}\approx\frac{4}{3}\mathbf{Q}^{2}+Q_{3}^{2}\left(\frac{1}{\sin^{2}{\gamma}}-\frac{4}{3}\right),
\label{app}
\end{equation}
where $\mathbf{Q}$ is the total angular momentum vector operator.

Assuming a factorized total wave function $\Psi(\beta,\gamma,\Omega)=\xi(\beta)\eta(\gamma)D^{L}_{MK}(\Omega)$ where $D^{L}_{MK}$ are Wigner functions of total angular momentum $L$ and its projections $M$ and $K$ on the body-fixed and laboratory-fixed $z$ axis respectively, the associated Schr\"{o}dinger equation is separated into $\beta$ and $\gamma$ parts:
\begin{eqnarray}
&&\left[-\frac{\partial^{2}}{\partial{\beta^{2}}}-\frac{2(2-\chi)}{\beta}\frac{\partial}{\partial{\beta}}+\frac{L(L+1)}{3\beta^{2}}+u(\beta)\right]\xi(\beta)\nonumber\\
&&=\epsilon_{\beta}\xi(\beta),\label{b}\\
&&(1-\chi)\left[-\frac{1}{\beta^{2}_{0}\sin{3\gamma}}\frac{\partial}{\partial\gamma}\sin{3\gamma}\frac{\partial}{\partial\gamma}\right.\nonumber\\
&&\left.+\frac{K^{2}}{4\beta^{2}_{0}}\left(\frac{1}{\sin^{2}{\gamma}}-\frac{4}{3}\right)+v(\gamma)\right]\eta(\gamma)=\epsilon_{\gamma}\eta(\gamma),\label{ecg}
\end{eqnarray}
where $\epsilon=\epsilon_{\beta}+\epsilon_{\gamma}=\frac{2B}{\hbar^{2}}E$. $\beta_{0}$ is a static "average" of $\beta$ which assures an approximated adiabatic separation of the $\beta$ and $\gamma$ surface oscillations. The advantages and shortcomings of this approximation were extensively analysed in Ref.\cite{Caprio}. The angular dependence was extracted through averaging on the Wigner states. The $\gamma$ equation is treated as in the usual $\gamma$-stable case \cite{Iachello1} by applying a harmonic approximation with respect to $\gamma=0$ for the involved trigonometric functions. The lowest order symmetry obeying $\gamma$ potential $v(\gamma)=a(1-\cos{3\gamma})$ gets the same treatment such that constant $a$ will acquire the role of the $\gamma$ oscillation stiffness. As a result, the $\gamma$ differential equation becomes:
\begin{eqnarray}
&&\left[-\frac{1}{\beta_{0}^{2}\gamma}\frac{\partial}{\partial\gamma}\gamma\frac{\partial}{\partial\gamma}+\left(\frac{K}{2}\right)^{2}\frac{1}{\beta_{0}^{2}\gamma^{2}}+(3a)^{2}\frac{\gamma^{2}}{2}\right]\eta(\gamma)\nonumber\\
&&=\epsilon'_{\gamma}\eta(\gamma),\,\,\,\,\epsilon'_{\gamma}=\frac{\epsilon_{\gamma}}{1-\chi}+\frac{K^{2}}{3\beta^{2}_{0}}.
\label{ecga}
\end{eqnarray}
Its similarity with the radial equation for a two-dimensional harmonic oscillator is obvious, such that one readily obtains the corresponding solutions as:
\begin{eqnarray}
\epsilon'_{\gamma}&=&\frac{3a}{\beta_{0}}(n_{\gamma}+1),\,\,n_{\gamma}=0,1,2,...,\label{eg}\\
\eta_{n_{\gamma}K}(\gamma)&=&N_{nK}\gamma^{\left|\frac{K}{2}\right|}e^{-3a\frac{\gamma^{2}}{2}}L_{n}^{|\frac{K}{2}|}(3a\gamma^{2}),\label{fg}
\end{eqnarray}
where $N_{nK}$ is a normalization constant, $n=(n_{\gamma}-|K|/2)/2$ with $K=0,\pm 2n_{\gamma}$ for $n_{\gamma}$ even and $K=\pm 2n_{\gamma}$ for $n_{\gamma}$ odd, respectively.

In accordance to $X(5)$ \cite{Iachello1} and $X(3)$ \cite{BonX3} critical point solutions, we will consider here an anharmonic behaviour reflected into a square well shape of the potential:
\begin{equation}
u(\beta)=
\left\{\begin{array}{l}
0,\,\,\beta\leqslant\beta_{W},\\
\infty,\,\,\beta>\beta_{W},
\end{array}\right.
\end{equation}
with $\beta_{W}$ indicating the position of the infinite wall. Making the change of variable $\xi(\beta)=\beta^{\chi-\frac{3}{2}}f(\beta)$, equation (\ref{b}) is written as a Bessel differential equation:
\begin{equation}
\left[\frac{\partial^{2}}{\partial{\beta^{2}}}+\frac{1}{\beta}\frac{\partial}{\partial{\beta}}+\left(k^{2}-\frac{\nu^{2}}{\beta^{2}}\right)\right]f(\beta)=0,
\end{equation}
where
\begin{equation}
\nu=\sqrt{\frac{L(L+1)}{3}+\left(\frac{3}{2}-\chi\right)^{2}}.
\end{equation}
The associated wave function must satisfy the boundary condition $f(\beta_{W})=0$, from which one extracts the $\beta$ energy spectrum in terms of the $s$-th zero $x_{s,\nu}$ of the Bessel function $J_{\nu}(x_{s,\nu}\beta/\beta_{W})$ \cite{Abram}:
\begin{equation}
\epsilon_{Ln_{\beta}}^{\beta}=\left(\frac{x_{n_{\beta}+1,\nu}}{\beta_{W}}\right)^{2}.
\end{equation}
At this point we assigned the $\beta$ vibration quantum number by $n_{\beta}=s-1$. Completing the $\beta$ eigensystem are the $\beta$ variable wave functions given as:
\begin{equation}
\xi_{Ln_{\beta}}(\beta)=N_{n_{\beta}\nu}\beta^{\chi-\frac{3}{2}}J_{\nu}(x_{n_{\beta}+1,\nu}\beta/\beta_{W}).
\end{equation}
$N_{n_{\beta}\nu}$ is the normalization constant which is computed using the properties of the Bessel functions:
\begin{eqnarray}
\left(N_{n_{\beta}\nu}\right)^{-2}&=&\int_{0}^{\beta_{W}}\beta\left[J_{\nu}(x_{n_{\beta}+1,\nu}\beta/\beta_{W})\right]^{2}d\beta\nonumber\\
&=&\frac{\beta_{W}^{2}}{2}\left[J_{\nu+1}(x_{n_{\beta}+1,\nu})\right]^{2}.
\end{eqnarray}
Note that in the above scalar product we used the modified integration measure which accounts for the shape phase mixing \cite{Noi2}.

Finally, the total excitation energy of the system in respect to the ground state is defined as
\begin{equation}
E_{LKn_{\beta}n_{\gamma}}=\frac{\hbar^{2}}{2B}\left[\epsilon^{\beta}_{Ln_{\beta}}+\epsilon_{Kn_{\gamma}}^{\gamma}-\epsilon^{\beta}_{00}-(1-\chi)\frac{3a}{\beta_{0}}\right].
\label{et}
\end{equation}
While the total solution of the Hamiltonian (\ref{Ht}) is given by the normalized and symmetrized product of angular, $\beta$ and $\gamma$ wave functions \cite{Iachello1,BonX5ES,BonDES}:
\begin{eqnarray}
&&\Psi_{LMKn_{\beta}n_{\gamma}}(\beta,\gamma,\Omega)=\xi_{Ln_{\beta}}(\beta)\eta_{n_{\gamma}|K|}(\gamma)\nonumber\\
&&\times\sqrt{\frac{2L+1}{16\pi^{2}(1+\delta_{K,0})}}\left[D_{MK}^{L}(\Omega)+(-)^{L}D_{M-K}^{L}(\Omega)\right].\nonumber\\
\end{eqnarray}

Transition rates can then be calculated by employing the general expression for the quadrupole transition operator,
\begin{equation}
T_{\mu}^{(E2)}=t\beta\left[D_{\mu0}^{2}\cos{\gamma}+\frac{1}{\sqrt{2}}\left(D_{\mu2}^{2}+D_{\mu-2}^{2}\right)\sin{\gamma}\right],
\label{TE2}
\end{equation}
where $t$ is a scaling factor. The transition probability can also be given in a factorized form as \cite{BonDES,Bijker}:
\begin{eqnarray}
B(E2,LKn_{\beta}n_{\gamma}\rightarrow L'K'n'_{\beta}n'_{\gamma})\nonumber\\
=\frac{5t^{2}}{16\pi}\left(C^{L\,\,2\,\,L'}_{KK'-KK'}B_{L'n'_{\beta}}^{Ln_{\beta}}G_{K'n'_{\gamma}}^{Kn_{\gamma}}\right)^{2}.
\end{eqnarray}
$G$ is the integral over the $\gamma$ shape variable of only the second term from (\ref{TE2}),
\begin{equation}
G_{K'n'_{\gamma}}^{Kn_{\gamma}}=\int_{0}^{\pi/3}\sin{\gamma}\eta_{n_{\gamma}K}\eta_{n'_{\gamma}K'}\left|\sin{3\gamma}\right|d\gamma,
\end{equation}
because in the present model $\gamma$ is very small and consequently $\cos{\gamma}\approx1$ \cite{BonDES}. $C$ is the Clebsch-Gordan coefficient dictating the angular momentum selection rules, while $B$ is defined as:
\begin{equation}
B_{L'n'_{\beta}}^{Ln_{\beta}}=\int_{0}^{\beta_{W}}\xi_{Ln_{\beta}}(\beta)\xi_{L'n'_{\beta}}(\beta)\beta^{5-2\chi}d\beta.
\label{bm}
\end{equation}

\begin{figure*}[th!]
\begin{center}
\includegraphics[clip,trim = 0mm 0mm 0mm 0mm,width=0.98\textwidth]{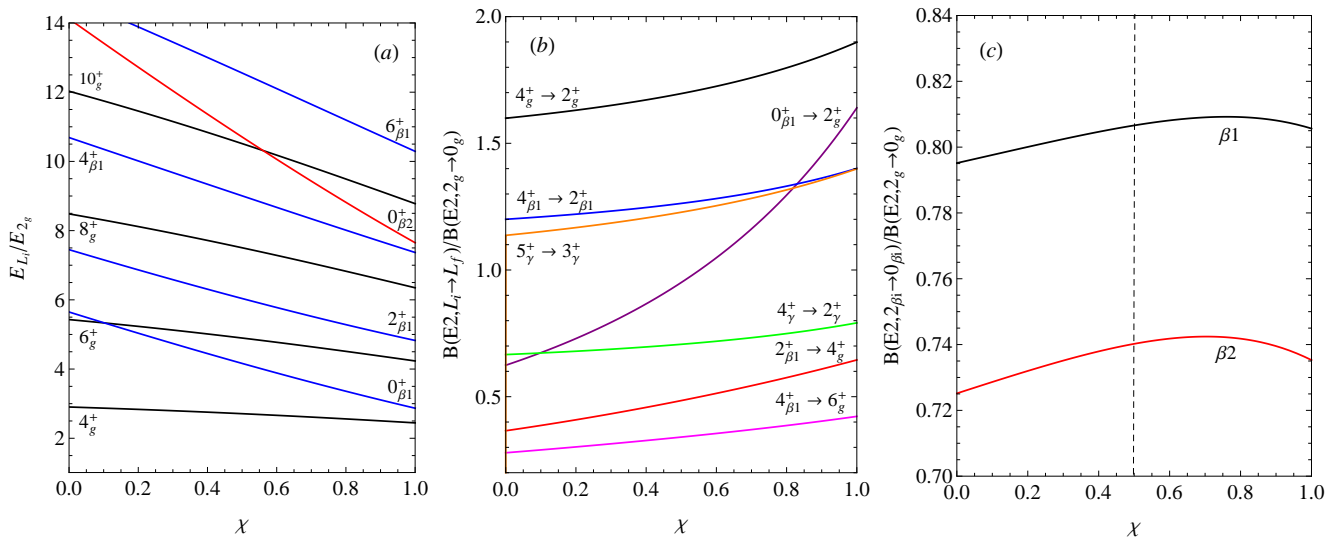}
\end{center}
\vspace{-0.2cm}
\caption{The low-lying energy spectrum of ground and first two $\beta$ excited bands (a) and few $\Delta K=0$ $B(E2)$ transition probabilities [(b) and (c)] are given as functions of the rigidity parameter $\chi$.}
\label{x3x5}
\end{figure*}

\renewcommand{\theequation}{4.\arabic{equation}}
\section{Numerical results}
\label{sec:3}

In order to find experimental counterparts of the shape phase space mixing, a thorough analysis is required on the evolution of its spectral properties as function of $\chi$. Besides $\chi$, the total energy function (\ref{et}) expressed in $\hbar^{2}/(2B)$ units, depends also on $a$ and $\beta_{0}$ through the $\gamma$ energy contribution. However, the excitation energies of the ground and $\beta$ band states are independent of $a$ and $\beta_{0}$, while the $\gamma$ band ($n_{\gamma}=1,K=2$) is shifted by the amount
\begin{equation}
\alpha=(1-\chi)\left(\frac{3a}{\beta_{0}}-\frac{4}{3\beta_{0}^{2}}\right).
\label{D}
\end{equation}
Therefore, we visualized in Fig.\ref{x3x5} the evolution of the energy spectrum corresponding to the ground and $\beta$ bands and few strong $\Delta K=0$ transition probabilities only as a function of $\chi$. The energy spectrum depicted in Fig.\ref{x3x5}(a) shows that the energy of all ground and $\beta$ states decreases linearly with $\chi$. Although not shown, the evolution of the even $L$ $\gamma$ band states follows those of the ground band states with a constant shift in energy. The slope of the energy level curves is almost the same for states belonging to the same vibrational band, with a variable increase with the energy of the state depending on the band. As a result, the steepest decrease in energy is associated to the highest vibrational quantum number. Due to different rates of energy decreasing, some levels belonging to different vibrational bands intersect each other at some values of the rigidity $\chi$. For example at $\chi=0.103$, $0_{\beta1}^{+}$ and $6^{+}_{g}$ states are degenerated, while the degeneracy of the $0_{\beta2}^{+}$ and $10^{+}_{g}$ states happens at $\chi=0.560$. The last intersection is very close to the case of $X(4)$ where the $\gamma$-rigid/stable mixing is equal. Regarding this special case, we can see that its $8_{g}^{+}$ state, which was shown in Ref.\cite{X4} that satisfies exactly the Euclidean dynamical symmetry $E(4)$, is positioned right in the middle between $2^{+}$ and $4^{+}$ states of the first excited $\beta$ band. For smaller values of $\chi$, the $8_{g}^{+}$ energy level shifts toward $2_{\beta1}^{+}$, and when $\chi$ is increased it gets closer to the $4_{\beta1}^{+}$ state. Although this observation is just a numerical peculiarity, it serves as another example of the median role played by the $X(4)$ model in the relation between $X(3)$ and $X(5)$.

As all states decrease in energy when $\chi$ increases, it is expected that the quadrupole transition probabilities would gain in value. Indeed, as can be seen from Fig.\ref{x3x5}(b), all $\Delta K=0$ transitions connecting the low lying states have greater probabilities for increased values of $\chi$. In contradistinction to the energy levels, the $B(E2)$ rates all have distinct mostly nonlinear evolution from $\chi=0$ to $\chi\to 1$. The transition most sensitive to $\chi$ variation is $0_{\beta2}^{+}\,\rightarrow\,2^{+}_{g}$, which might be taken as a distinguishable observable instead of the purely theoretical rigidity parameter. On the other hand, the most insensitive transitions are the first inband transitions from the two $\beta$ bands. Moreover, their evolution with $\chi$ shown in Fig.\ref{x3x5}(c) is not even monotonous, but has maximum points in the region of $\chi=0.7$.

The model is applied to $X(5)$ and $X(3)$ nuclei as well as to their isotopic and isotonic neighbours. The experimental ground, $\gamma$ and available $\beta$ band energies normalized to the excitation energy of the $2_{g}^{+}$ state are fitted against rigidity $\chi$ which completely describes the ground and $\beta$ excited bands, and the parameter $\alpha$ (\ref{D}) fixing the $\gamma$ band energy shift in respect to the ground state. The fitness of the present theoretical description is judged by the standard error
\begin{equation}
\sigma=\sqrt{\frac{1}{N-1}\sum_{i=1}^{N}\left[\frac{E_{i}(Th)}{E_{2_{g}^{+}}(Th)}-\frac{E_{i}(Exp)}{E_{2_{g}^{+}}(Exp)}\right]^{2}}.
\label{sigma}
\end{equation}
As the minimization of the above quantity tends to be disadvantageous for the low energy states, we consider for the fitting procedure for the ground band only experimental data up to $L=26$. Such a restriction assures an overall realistic fit of the involved free parameters.

There are two regions of the nuclide chart where such critical phenomena are expected. The first is the set of rare earth nuclei around $N=90$ where the $X(5)$ behaviour was originally found. The other domain is localized around $N=100$ and consists of Os and Pt isotopes where $X(3)$ candidates were pointed out \cite{BonX3} and a second island of $X(5)$ experimental realisation was predicted \cite{Dewald}. In what follows we will present the results of the fits in part for each of these groups of nuclei.

\setlength{\tabcolsep}{5.5pt}
\begin{table*}[ht!]
\caption{Theoretical results for ground, $\gamma$ and first two $\beta$ band energies normalized to the energy of the first excited state $2_{g}^{+}$ are compared with the available experimental data for $^{148}$Ce\cite{148Ce}, $^{150}$Nd\cite{150Nd}, $^{156}$Dy\cite{156Dy} and $^{158}$Er\cite{158Er}. The dimensionless parameters $\chi$ and $\alpha$ are also given together with the corresponding deviation $\sigma$ defined by (\ref{sigma}). Values in parentheses denote states with uncertain assignment of angular momentum and therefore were excluded from the fits.}
\label{tab:1}
{\footnotesize
\begin{center}
\begin{tabular}{ccccccccc}
\hline\hline\noalign{\smallskip}
&\multicolumn{2}{c}{$^{148}$Ce}&\multicolumn{2}{c}{$^{150}$Nd}&\multicolumn{2}{c}{$^{156}$Dy}&\multicolumn{2}{c}{$^{158}$Er}\\
\noalign{\smallskip}\hline\noalign{\smallskip}
$L$&~~Exp.~~&~~Th.~~&~~Exp.~~&Th.~~&~~Exp.~~&~~Th.~~&~~Exp.~~&~~Th.~~\\
\noalign{\smallskip}\hline\noalign{\smallskip}
 $2_{g}^{+}$      &  1.00 &  1.00 &  1.00 &  1.00 &  1.00 &  1.00 &  1.00 &  1.00\\
 $4_{g}^{+}$      &  2.86 &  2.66 &  2.93 &  2.85 &  2.93 &  2.81 &  2.74 &  2.61\\
 $6_{g}^{+}$      &  5.30 &  4.77 &  5.53 &  5.29 &  5.59 &  5.15 &  5.05 &  4.65\\
 $8_{g}^{+}$      &  8.14 &  7.27 &  8.68 &  8.22 &  8.82 &  7.97 &  7.77 &  7.07\\
 $10_{g}^{+}$     & 11.30 & 10.16 & 12.28 & 11.62 & 12.52 & 11.22 & 10.79 &  9.85\\
 $12_{g}^{+}$     & 14.69 & 13.43 & 16.27 & 15.46 & 16.59 & 14.91 & 13.95 & 13.00\\
 $14_{g}^{+}$     & 18.22 & 17.06 & 20.59 & 19.74 & 20.96 & 19.01 & 17.56 & 16.49\\
 $16_{g}^{+}$     & 21.86 & 21.05 & 25.19 & 24.44 & 25.57 & 23.52 & 20.95 & 20.34\\
 $18_{g}^{+}$     & 25.66 & 25.39 &       & 29.58 & 30.33 & 28.43 & 24.32 & 24.52\\
 $20_{g}^{+}$     & 29.57 & 30.09 &       & 35.13 & 35.27 & 33.75 & 27.96 & 29.05\\
 $22_{g}^{+}$     & 33.52 & 35.15 &       & 41.10 & 40.45 & 39.46 &       & 33.91\\
 $24_{g}^{+}$     &       & 40.55 &       & 47.48 & 45.94 & 45.57 &       & 39.11\\
 $26_{g}^{+}$     &       & 46.29 &       & 54.27 & 51.76 & 52.07 &       & 44.65\\
\noalign{\smallskip}\hline\noalign{\smallskip}
 $0_{\beta1}^{+}$ &  4.86 &  3.87 &  5.19 &  5.20 &  4.90 &  4.81 &  4.20 &  3.63\\
 $2_{\beta1}^{+}$ &  5.90 &  5.77 &  6.53 &  7.02 &  6.01 &  6.65 &  5.15 &  5.54\\
 $4_{\beta1}^{+}$ &  7.72 &  8.66 &  8.74 & 10.20 &  7.90 &  9.76 &  6.54 &  8.35\\
 $6_{\beta1}^{+}$ &       & 12.08 & 11.83 & 14.12 & 10.43 & 13.55 &       & 11.67\\
 $8_{\beta1}^{+}$ &       & 15.96 &       & 18.63 & 13.49 & 17.89 &       & 15.41\\
$10_{\beta1}^{+}$ &       & 20.25 &       & 23.66 & 16.81 & 22.72 &       & 19.55\\
 \noalign{\smallskip}\hline\noalign{\smallskip}
 $0_{\beta2}^{+}$ &       & 10.03 &(13.35)& 13.10 &(10.00)& 12.20 & (7.22)&  9.46\\
 $2_{\beta2}^{+}$ &       & 12.81 &       & 15.74 &       & 14.88 &       & 12.27\\
\noalign{\smallskip}\hline\noalign{\smallskip}
 $2_{\gamma}^{+}$ & (6.25)&  6.24 &  8.16 &  8.35 &  6.46 &  7.03 &  4.27 &  4.59\\
 $3_{\gamma}^{+}$ &  7.05 &  7.01 &  9.22 &  9.20 &  7.42 &  7.86 &  5.43 &  5.34\\
 $4_{\gamma}^{+}$ &       &  7.91 & 10.39 & 10.21 &  8.48 &  8.84 &  6.16 &  6.20\\
 $5_{\gamma}^{+}$ &  8.98 &  8.91 &       & 11.36 &  9.69 &  9.95 &  7.48 &  7.17\\
 $6_{\gamma}^{+}$ &       & 10.01 &       & 12.64 & 11.07 & 11.18 &  8.27 &  8.24\\
 $7_{\gamma}^{+}$ & 11.27 & 11.22 &       & 14.05 & 12.55 & 12.53 &  9.96 &  9.40\\
 $8_{\gamma}^{+}$ &       & 12.52 &       & 15.57 & 14.22 & 14.00 & 10.51 & 10.65\\
 $9_{\gamma}^{+}$ & 13.88 & 13.92 &       & 17.22 & 15.91 & 15.57 &       & 12.00\\
$10_{\gamma}^{+}$ &       & 15.41 &       & 18.97 & 17.77 & 17.26 & 12.95 & 13.44\\
$11_{\gamma}^{+}$ & 16.87 & 16.99 &       & 20.84 & 19.69 & 19.04 &       & 14.97\\
\noalign{\smallskip}\hline\noalign{\smallskip}
$\chi$ &\multicolumn{2}{c}{0.605}&\multicolumn{2}{c}{0.145}&\multicolumn{2}{c}{0.276} &\multicolumn{2}{c}{0.696}\\
$\alpha$ &\multicolumn{2}{c}{45.36}&\multicolumn{2}{c}{53.80}&\multicolumn{2}{c}{46.09} &\multicolumn{2}{c}{32.21}\\
Nr. states  &\multicolumn{2}{c}{18}&\multicolumn{2}{c}{14}&\multicolumn{2}{c}{28}&\multicolumn{2}{c}{20}\\
$\sigma$  &\multicolumn{2}{c}{0.777}&\multicolumn{2}{c}{0.859}&\multicolumn{2}{c}{1.794}&\multicolumn{2}{c}{0.639}\\
\noalign{\smallskip}\hline\hline
\end{tabular}
\end{center}}
\vspace{-0.5cm}
\end{table*}

\setlength{\tabcolsep}{5.5pt}
\begin{table*}[ht!]
\caption{Same as in Table \ref{tab:1} but for $^{174}$Os\cite{174Os}, $^{176}$Os\cite{176Os,Os176g26}, $^{178}$Os\cite{178Os} and $^{180}$Os\cite{180OsPt}.}
\label{tab:2}
{\footnotesize
\begin{center}
\begin{tabular}{ccccccccc}
\hline\hline\noalign{\smallskip}
&\multicolumn{2}{c}{$^{174}$Os}&\multicolumn{2}{c}{$^{176}$Os}&\multicolumn{2}{c}{$^{178}$Os}&\multicolumn{2}{c}{$^{180}$Os}\\
\noalign{\smallskip}\hline\noalign{\smallskip}
$L$&~~Exp.~~&Th.~~&~~Exp.~~&~~Th.~~&~~Exp.~~&~~Th.~~&~~Exp.~~&~~Th.~~\\
\noalign{\smallskip}\hline\noalign{\smallskip}
 $2_{g}^{+}$      &  1.00 &  1.00 &  1.00 &  1.00 &  1.00 &  1.00 &  1.00 &  1.00\\
 $4_{g}^{+}$      &  2.74 &  2.57 &  2.93 &  2.82 &  3.01 &  2.89 &  3.09 &  2.88\\
 $6_{g}^{+}$      &  4.90 &  4.54 &  5.50 &  5.20 &  5.76 &  5.39 &  6.02 &  5.35\\
 $8_{g}^{+}$      &  7.39 &  6.87 &  8.57 &  8.05 &  9.04 &  8.40 &  9.52 &  8.34\\
 $10_{g}^{+}$     & 10.20 &  9.56 & 12.10 & 11.35 & 12.73 & 11.90 & 13.38 & 11.81\\
 $12_{g}^{+}$     & 13.33 & 12.59 & 16.05 & 15.08 & 16.81 & 15.86 & 17.48 & 15.73\\
 $14_{g}^{+}$     & 16.75 & 15.96 & 20.39 & 19.24 & 21.24 & 20.27 & 21.76 & 20.10\\
 $16_{g}^{+}$     & 20.43 & 19.66 & 25.03 & 23.81 & 25.97 & 25.13 & 26.45 & 24.90\\
 $18_{g}^{+}$     & 24.35 & 23.69 & 29.75 & 28.80 & 31.36 & 30.42 & 31.30 & 30.14\\
 $20_{g}^{+}$     & 28.53 & 28.06 & 34.67 & 34.19 & 36.83 & 36.14 & 36.50 & 35.81\\
 $22_{g}^{+}$     & 32.99 & 32.74 & 39.96 & 39.98 & 42.30 & 42.30 & 42.02 & 41.91\\
 $24_{g}^{+}$     & 37.75 & 37.75 & 45.50 & 46.18 & 48.61 & 48.88 & 47.87 & 48.42\\
 $26_{g}^{+}$     & 42.79 & 43.09 &(51.54)& 52.77 &       & 55.89 & 54.08 & 55.36\\
\noalign{\smallskip}\hline\noalign{\smallskip}
 $0_{\beta1}^{+}$ &  3.44 &  3.41 &  4.45 &  4.93 &  4.93 &  5.50 &  5.57 &  5.40\\
 $2_{\beta1}^{+}$ & (4.36)&  5.33 &  5.50 &  6.77 &  5.84 &  7.31 &  6.29 &  7.22\\
 $4_{\beta1}^{+}$ &  6.24 &  8.07 &  7.59 &  9.90 &  7.75 & 10.53 &  7.97 & 10.42\\
 $6_{\beta1}^{+}$ &  8.98 & 11.28 & 10.60 & 13.73 & 10.58 & 14.55 & 10.44 & 14.41\\
 $8_{\beta1}^{+}$ &       & 14.89 &       & 18.13 &       & 19.19 &       & 19.01\\
 \noalign{\smallskip}\hline\noalign{\smallskip}
 $0_{\beta2}^{+}$ &       &  8.93 &       & 12.48 &       & 13.79 &       & 13.56\\
 $2_{\beta2}^{+}$ &       & 11.77 &       & 15.15 &       & 16.40 &       & 16.18\\
\noalign{\smallskip}\hline\noalign{\smallskip}
 $2_{\gamma}^{+}$ &  5.34 &  6.06 &  6.39 &  6.99 &  6.54 &  8.01 &  6.59 &  7.70\\
 $3_{\gamma}^{+}$ &  6.64 &  6.79 &  7.68 &  7.82 &  7.81 &  8.87 &  7.74 &  8.55\\
 $4_{\gamma}^{+}$ &  7.91 &  7.63 &  9.06 &  8.81 &  9.19 &  9.90 &  9.06 &  9.58\\
 $5_{\gamma}^{+}$ &  9.16 &  8.57 & 10.43 &  9.94 & 10.83 & 11.08 & 10.64 & 10.75\\
 $6_{\gamma}^{+}$ &       &  9.59 &       & 11.19 & 14.87 & 12.40 & 12.32 & 12.05\\
 $7_{\gamma}^{+}$ &       & 10.72 &       & 12.56 & 14.86 & 13.84 & 14.24 & 13.49\\
 $8_{\gamma}^{+}$ &       & 11.93 &       & 14.04 &       & 15.41 &       & 15.04\\
 $9_{\gamma}^{+}$ &       & 13.23 &       & 15.64 &       & 17.10 & 18.25 & 16.72\\
\noalign{\smallskip}\hline\noalign{\smallskip}
$\chi$&\multicolumn{2}{c}{0.781} &\multicolumn{2}{c}{0.235}&\multicolumn{2}{c}{0.046}&\multicolumn{2}{c}{0.079}\\
$\alpha$&\multicolumn{2}{c}{47.19} &\multicolumn{2}{c}{45.15}&\multicolumn{2}{c}{49.70}&\multicolumn{2}{c}{47.98}\\
Nr. states  &\multicolumn{2}{c}{19}&\multicolumn{2}{c}{19}&\multicolumn{2}{c}{21}&\multicolumn{2}{c}{23}\\
$\sigma$  &\multicolumn{2}{c}{1.674}&\multicolumn{2}{c}{1.122}&\multicolumn{2}{c}{1.419}&\multicolumn{2}{c}{1.386}\\
\noalign{\smallskip}\hline\hline
\end{tabular}
\end{center}}
\vspace{-0.5cm}
\end{table*}

\setlength{\tabcolsep}{5.5pt}
\begin{table*}[ht!]
\caption{Same as in Table \ref{tab:1} but for $^{178}$Pt\cite{178Pt}, $^{180}$Pt\cite{180OsPt}, $^{182}$Pt\cite{182Pt} and $^{184}$Pt\cite{184Pt}}
\label{tab:3}
{\footnotesize
\begin{center}
\begin{tabular}{ccccccccc}
\hline\hline\noalign{\smallskip}
&\multicolumn{2}{c}{$^{178}$Pt}&\multicolumn{2}{c}{$^{180}$Pt}&\multicolumn{2}{c}{$^{182}$Pt}&\multicolumn{2}{c}{$^{184}$Pt}\\
\noalign{\smallskip}\hline\noalign{\smallskip}
$L$&~~Exp.~~&~~Th.~~&~~Exp.~~&Th.~~&~~Exp.~~&~~Th.~~&~~Exp.~~&~~Th.~~\\
\noalign{\smallskip}\hline\noalign{\smallskip}
 $2_{g}^{+}$     &  1.00 &  1.00 &  1.00 &  1.00 &  1.00 &  1.00 &  1.00 &  1.00 \\
 $4_{g}^{+}$     &  2.51 &  2.54 &  2.68 &  2.73 &  2.71 &  2.65 &  2.67 &  2.54 \\
 $6_{g}^{+}$     &  4.49 &  4.47 &  4.94 &  4.96 &  5.00 &  4.73 &  4.90 &  4.46 \\
 $8_{g}^{+}$     &  6.92 &  6.75 &  7.71 &  7.62 &  7.78 &  7.21 &  7.55 &  6.74 \\
 $10_{g}^{+}$    &  9.76 &  9.37 & 10.93 & 10.69 & 10.96 & 10.07 & 10.47 &  9.36 \\
 $12_{g}^{+}$    & 12.97 & 12.34 & 14.55 & 14.15 & 14.47 & 13.30 & 13.53 & 12.32 \\
 $14_{g}^{+}$    & 16.52 & 15.63 & 18.55 & 18.01 & 18.27 & 16.89 & 16.73 & 15.61 \\
 $16_{g}^{+}$    & 20.31 & 19.25 & 22.88 & 22.25 & 22.33 & 20.83 & 20.14 & 19.22 \\
 $18_{g}^{+}$    & 24.13 & 23.19 & 27.76 & 26.87 & 26.42 & 25.13 & 23.74 & 23.16 \\
 $20_{g}^{+}$    & 27.91 & 27.45 & 32.54 & 31.87 & 30.51 & 29.78 & 27.57 & 27.42 \\
 $22_{g}^{+}$    & 31.89 & 32.03 & 37.39 & 37.24 & 34.87 & 34.78 & 31.70 & 31.99 \\
 $24_{g}^{+}$    &(36.17)& 36.93 & 42.76 & 42.99 & 39.54 & 40.12 & 36.18 & 36.88 \\
 $26_{g}^{+}$    &       & 42.14 & 48.52 & 49.10 & 44.56 & 45.80 & 41.02 & 42.09 \\
\noalign{\smallskip}\hline\noalign{\smallskip}
 $0_{\beta1}^{+}$&  2.47 &  3.27 &  3.12 &  4.31 &  3.22 &  3.80 &  3.02 &  3.27 \\
 $2_{\beta1}^{+}$&  3.84 &  5.20 &  5.62 &  6.18 &  5.53 &  5.70 &  5.18 &  5.20 \\
 $4_{\beta1}^{+}$&  6.21 &  7.90 &  8.15 &  9.18 &  8.00 &  8.57 &  7.57 &  7.90 \\
 $6_{\beta1}^{+}$&  8.67 & 11.05 & 10.77 & 12.79 & 10.64 & 11.96 & 11.04 & 11.03 \\
 $8_{\beta1}^{+}$& 12.90 & 14.58 &       & 16.89 & 13.66 & 15.79 &       & 14.56 \\
 \noalign{\smallskip}\hline\noalign{\smallskip}
 $0_{\beta2}^{+}$&       &  8.62 & (7.69)& 11.05 & (7.43)&  9.86 &       &  8.60 \\
 $2_{\beta2}^{+}$&       & 11.48 &       & 13.78 &       & 12.65 &       & 11.46 \\
\noalign{\smallskip}\hline\noalign{\smallskip}
 $2_{\gamma}^{+}$&       &  5.16 &  4.42 &  5.51 &  4.31 &  5.37 &  3.98 &  5.16\\
 $3_{\gamma}^{+}$& (5.88)&  5.88 &  6.28 &  6.32 &  6.08 &  6.13 &  5.77 &  5.87\\
 $4_{\gamma}^{+}$&       &  6.70 &  6.85 &  7.25 &  6.67 &  7.02 &  6.31 &  6.69\\
 $5_{\gamma}^{+}$&       &  7.62 &  8.58 &  8.31 &  8.42 &  8.01 &  8.02 &  7.61\\
 $6_{\gamma}^{+}$&       &  8.63 &       &  9.48 &  9.28 &  9.10 &  8.97 &  8.62\\
 $7_{\gamma}^{+}$&       &  9.72 & 11.27 & 10.75 & 11.17 & 10.30 & 10.62 &  9.71\\
 $8_{\gamma}^{+}$&       & 10.91 &       & 12.13 &       & 11.58 &       & 10.90\\
 $9_{\gamma}^{+}$&       & 12.18 & 14.35 & 13.62 &       & 12.97 &       & 12.52\\
\noalign{\smallskip}\hline\noalign{\smallskip}
$\chi$&\multicolumn{2}{c}{0.833} &\multicolumn{2}{c}{0.448} &\multicolumn{2}{c}{0.632}&\multicolumn{2}{c}{0.836}\\
$\alpha$&\multicolumn{2}{c}{39.71} &\multicolumn{2}{c}{36.73} &\multicolumn{2}{c}{38.24}&\multicolumn{2}{c}{39.70}\\
Nr. states  &\multicolumn{2}{c}{15}&\multicolumn{2}{c}{22}&\multicolumn{2}{c}{23}&\multicolumn{2}{c}{22}\\
$\sigma$  &\multicolumn{2}{c}{1.079}&\multicolumn{2}{c}{0.728}&\multicolumn{2}{c}{0.931}&\multicolumn{2}{c}{0.691}\\
\noalign{\smallskip}\hline\hline
\end{tabular}
\end{center}}
\vspace{-0.5cm}
\end{table*}

\begin{figure*}[th!]
\begin{center}
\includegraphics[clip,trim = 0mm 0mm 0mm 0mm,width=0.88\textwidth]{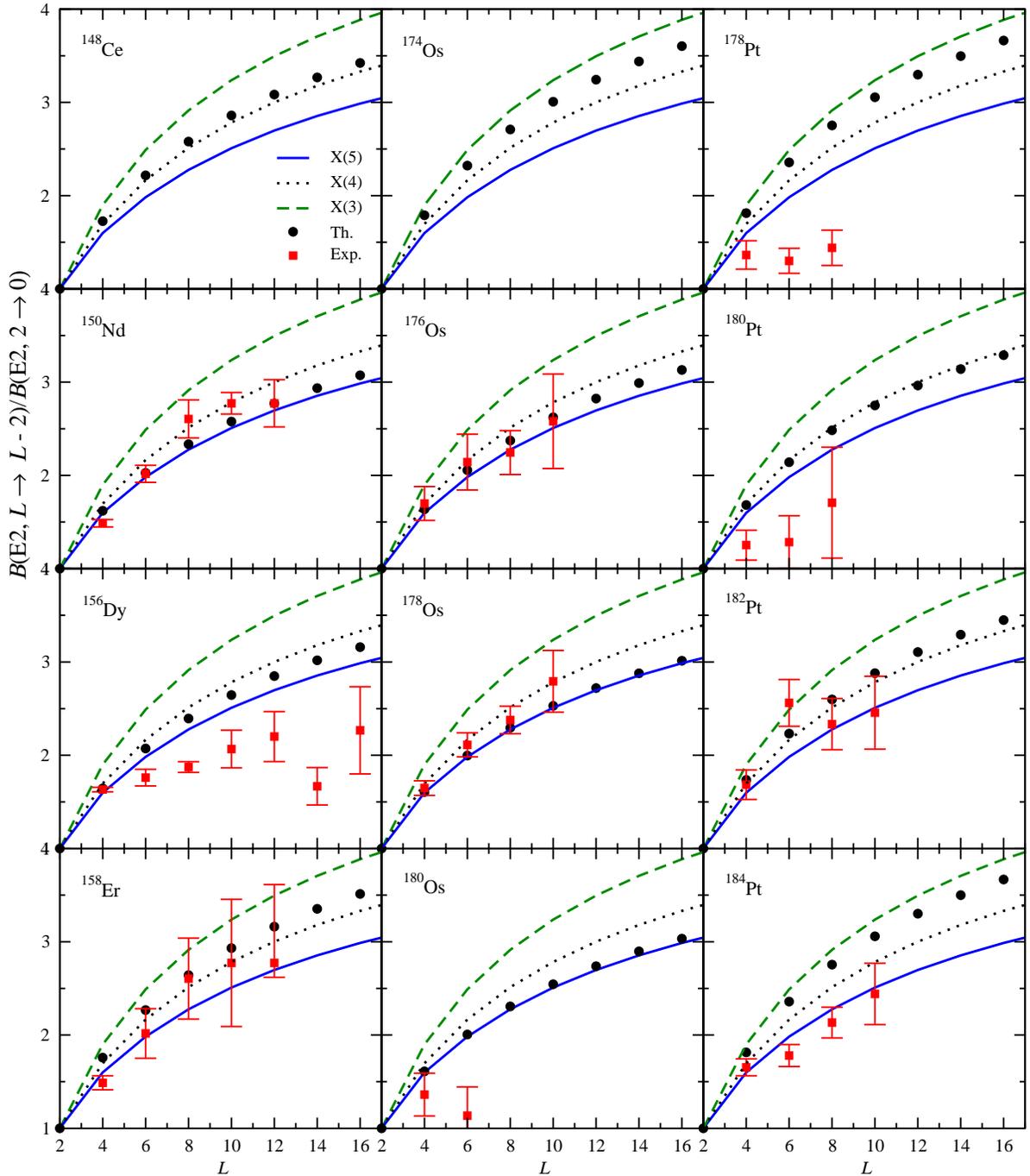}
\end{center}
\vspace{-0.2cm}
\caption{Theoretical ground state to ground state $E2$ transition probabilities normalized to the $2_{g}^{+}\rightarrow0^{+}_{g}$ transition are compared with the available experimental data corresponding to all considered nuclei and with the $X(3)$ \cite{BonX3}, $X(4)$ \cite{X4} and $X(5)$ \cite{Iachello1} predictions. All the data are gathered from Nuclear Data Sheets \cite{148Ce,150Nd,156Dy,158Er,174Os,176Os,178Os,180OsPt,182Pt,184Pt}, with the exception of $^{176,178}$Os and $^{180}$Pt nuclei, whose experimental values are extracted from Refs.\cite{Melon,Dewald} and respectively \cite{Pt180t}. Also the $E2$ rate for the $2_{g}^{+}\rightarrow0^{+}_{g}$ transition of $^{178}$Pt was taken from Ref.\cite{Pt178ge2}.}
\label{tg}
\end{figure*}

\begin{figure}[t!]
\begin{center}
\includegraphics[clip,trim = 0mm 0mm 0mm 0mm,width=0.46\textwidth]{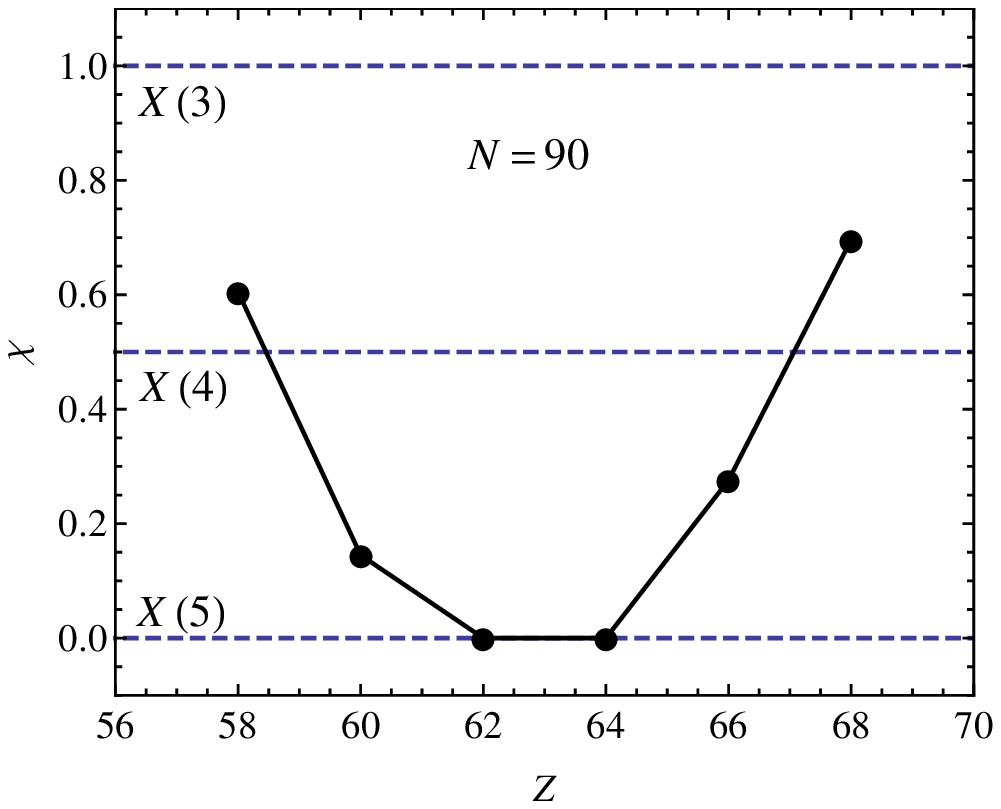}
\end{center}
\vspace{-0.2cm}
\caption{The fitted values of $\chi$ from Table \ref{tab:1} for the $N=90$ nuclei as well as the two $\chi=0$ values obtained for other two $N=90$ $X(5)$ nuclei are plotted as a function of atomic number $Z$.}
\label{N}
\end{figure}

\subsection{$N=90$ nuclei}

The best results with non-vanishing $\gamma$ rigidity were obtained in the $N=90$ region for $^{148}$Ce, $^{150}$Nd, $^{156}$Dy and $^{158}$Er. We can see that the selected nuclei encompass other two $N=90$ nuclei, $^{152}$Sm and $^{154}$Gd, which along with $^{150}$Nd \cite{150Ndc} and $^{156}$Dy \cite{156Dyc} are another well known $X(5)$ candidates \cite{Iachello1,154Gdc}. As a matter of fact, the fitting of their experimental energy spectra within the present model, provided $\chi=0$, a fact which confirms their complete $\gamma$-stable softness. Although not considered as $X(5)$ representatives, the criticality of the marginal $N=90$ isotopes, $^{148}$Ce and $^{158}$Er nuclei, is also well known \cite{Yu1,Yu2,Bon03}. The comparison of the theoretical and experimental energy spectra made in Table \ref{tab:1} shows that the best agreement is obtained especially for these two nuclei. Their deviation from $X(5)$ symmetry is also supported by their high values of $\chi$ which shows a more accentuated $\gamma$-rigid structure. Although the $\beta$ band states are the main source of discrepancies for all four $N=90$ nuclei, the model predicts quite well the position of the first and even second $\beta$ bandhead state for $^{150}$Nd and $^{156}$Dy. Especially good reproduction is obtained in all cases for the $\gamma$ band states. In what concerns the ground to ground $E2$ transition rates compared in Fig.\ref{tg}, the non vanishing $\chi$ values obtained in the fits for $^{150}$Nd and $^{158}$Er offered an excellent agreement with experimental data. While the poorer agreement with experimental spectrum for $^{156}$Dy is perpetuated also to the $E2$ transition probabilities. Looking once again in Table \ref{tab:1} at the fitted $\chi$ values, one observes that there is a regular evolution of the $\gamma$ rigidity along the $N=90$ nuclei. As can be seen from Fig.\ref{N}, the maximal $\gamma$-stable softness is obviously at the two purely $X(5)$ nuclei, $^{152}$Sm and $^{154}$Gd, which starts to abate in both directions towards $^{148}$Ce and $^{158}$Er nuclei, exhibiting the highest $\chi$ values. The $\chi$ values of the latter two nuclei are not exceedingly high but actually just above the $\chi=0.5$ value, a fact which indicated these nuclei as possible candidates of the $X(4)$ symmetry \cite{X4}.

\begin{figure}[t!]
\begin{center}
\includegraphics[clip,trim = 0mm 0mm 0mm 0mm,width=0.46\textwidth]{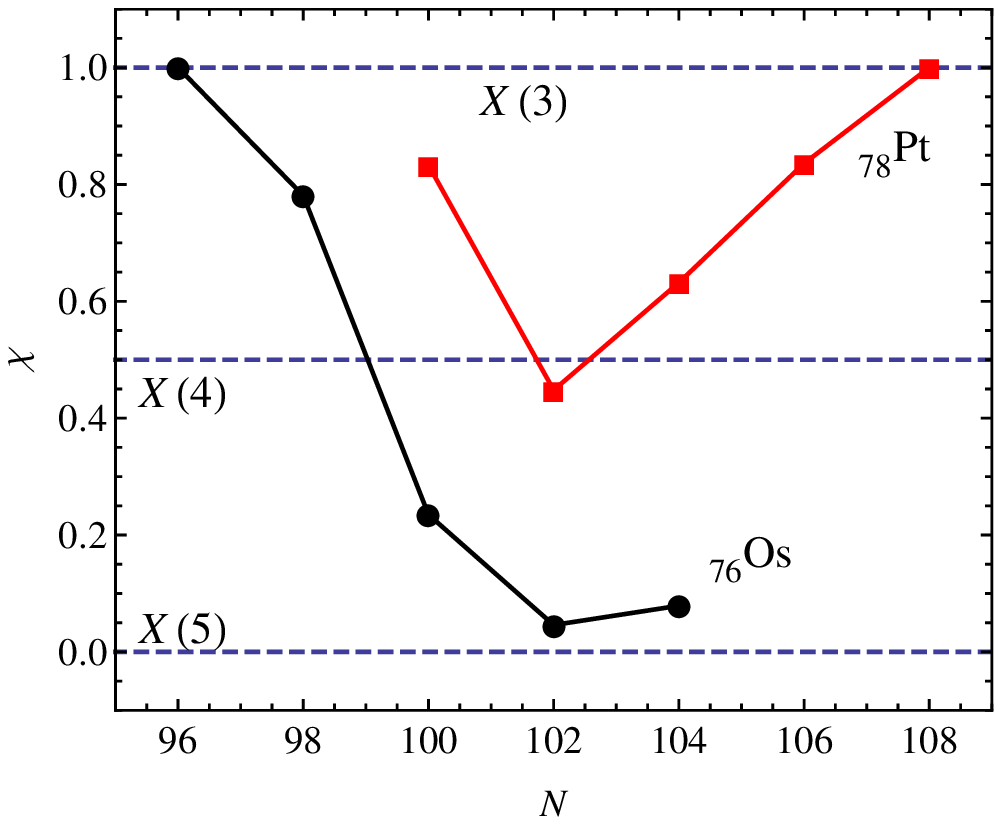}
\end{center}
\vspace{-0.2cm}
\caption{The fitted values of $\chi$ from Tables \ref{tab:2} and \ref{tab:3} for the Os and Pt isotopes as well as the two $\chi=1$ values corresponding to the $X(3)$ nuclei $^{172}$Os and $^{186}$Pt are plotted as functions of neutron number $N$.}
\label{Z}
\end{figure}

\begin{figure}[th!]
\begin{center}
\includegraphics[clip,trim = 0mm 0mm 0mm 0mm,width=0.44\textwidth]{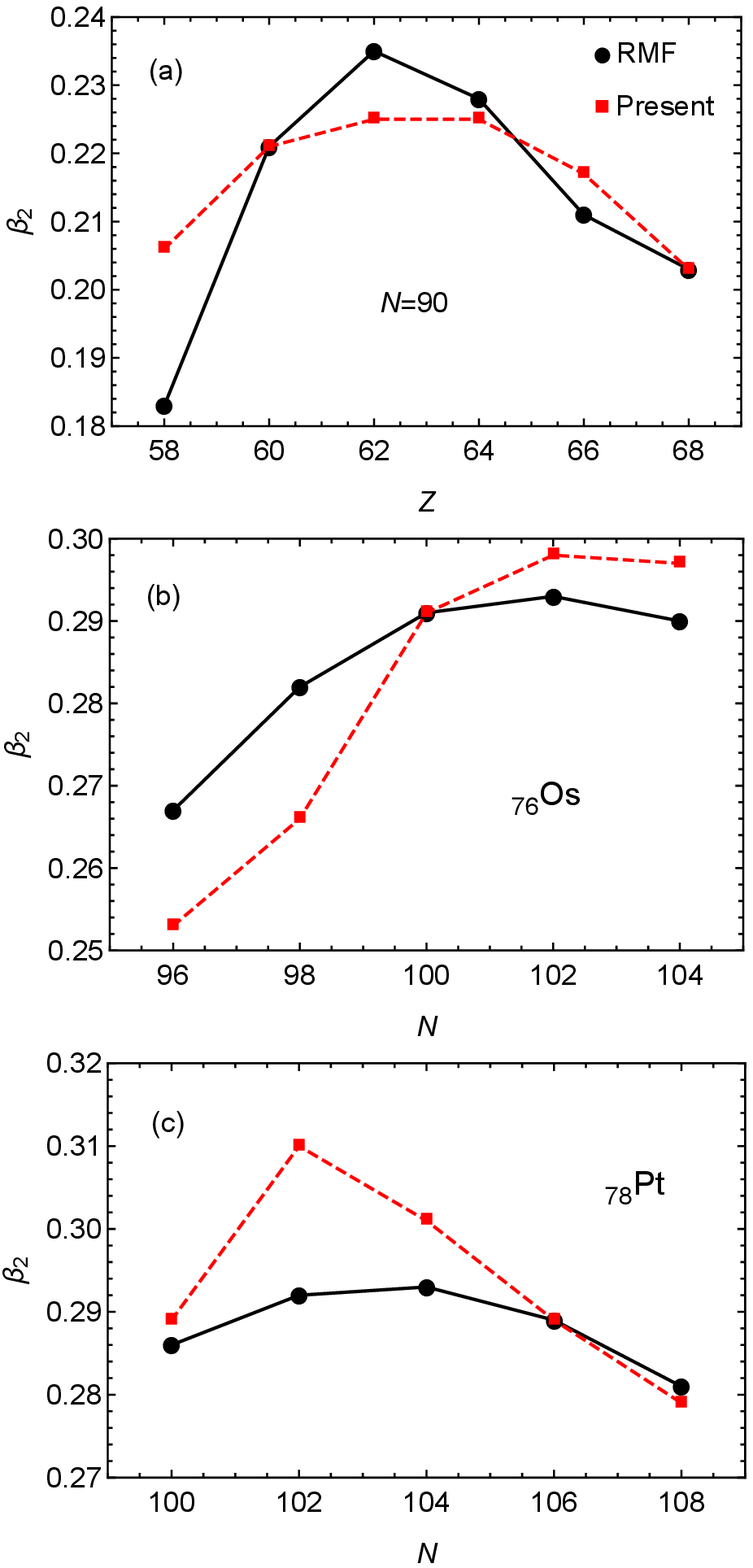}
\end{center}
\vspace{-0.2cm}
\caption{The quadrupole deformation $\beta_{2}$ calculated with RMF and the ground state average of $\beta$ (\ref{bm}) scaled such that to reproduce the RMF value for the nucleus with the best rms value, is given as function of $Z$ for $N=90$ nuclei (a) and as function of $N$ for Os (b) and Pt (c) isotopes.}
\label{b2}
\end{figure}

\subsection{Os and Pt isotopes}

The $A=180$ region provided similarly good results from fits of the $\gamma$ rigidity. The rms values from Table II are, however, lower than those reported in Table III for the Pt isotopes. Once again, the first $\beta$ band head is well reproduced, with exceptional matches in the cases of $^{174,180}$Os and $^{184}$Pt. The major distinction from the results of the $N=90$ nuclei is the larger deviations from experimental $\gamma$ band states, which is ascribed to the possible triaxial deformation of these nuclei. Also, while the values of parameter $\alpha$ are highly variate in $N=90$ isotones, for Os and Pt isotopes these are almost constant for each chain, with a lower average in the case of Pt nuclei. This is in agreement with the conspicuously constant structure of the observed low energy spectra for these nuclei \cite{McC1,McC2,Garcia1,Garcia2}. Moreover, their isotopic trajectory in the IBM symmetry triangle was found to be quiet concentrated and positioned centrally near the shape phase transition region \cite{McC1,McC2}. This aspect shows why these nuclei are found as experimental realisations of very different theoretical approaches.

Studying now the obtained $\chi$ values, we can see that the Os isotopes, with an exception, prefer small rigidity, while all considered Pt nuclei are highly $\gamma$-rigid. The lowest $\chi$ values for Pt isotopes are obtained for $^{180}$Pt and $^{182}$Pt, which are near the $\chi=0.5$ instance of the $X(4)$ model. It is then not surprising that these nuclei were considered as good experimental realizations of the $X(4)$ model \cite{X4}. On the other hand, the small values of $\chi$ obtained for $^{176,178,180}$Os supports their candidacy for the $X(5)$ model proposed in Ref.\cite{Dewald}. $^{176}$Os nucleus have however a non-negligible shape phase mixing. Moreover, its lighter neighbour have an even higher $\gamma$ rigidity. This ascending trend of $\gamma$ rigidity with decreasing neutron number $N$ shown in Fig.\ref{Z} continues to the fully $\gamma$-rigid nucleus $^{172}$Os which is one of the best $X(3)$ model candidates \cite{BonX3}. It is worth mentioning that a similar fitting of the $^{172}$Os energy spectrum confirmed its complete rigidity. The same is true for the other $X(3)$ candidate nucleus, $^{186}$Pt, from which $\gamma$ rigidity subsides when $N$ decreases. As can be seen from the same Fig.\ref{Z}, this evolution is even smoother than in the Os nuclei, being almost linear, but does not continue onto $^{178}$Pt. Although, its fitting results are satisfactory, the latter nucleus completely falls out of this trend with a very high $\gamma$ rigidity. This change, makes the $^{180}$Pt nucleus a singular or a terminal point in the evolution of $\gamma$ rigidity throughout this specific isotopic interval. A similar but less striking minimum is observed in the $^{178}$Os nucleus with respect to its isotopic chain, which has a slightly smaller $\chi$ value in comparison to both its neighbouring isotopes. Moreover, both critical nuclei, $^{180}$Pt and $^{178}$Os, have the same neutron number $N=102$ and their $\chi$ values are the closest realizations of $X(4)$ and $X(5)$ limits, respectively. While the evolution of $\gamma$ rigidity in Os isotopes is in agreement with the IBM results of Ref.\cite{McC1}, the increase in $\gamma$ rigidity for Pt isotopes is in contradiction with recent studies and the classical conception about heavier Pt isotopes as good $O(6)$ realisations. Indeed, the same interval of Pt isotopes exhibits an increasing $\gamma$-unstable softness for higher neutron numbers in the results obtained within the self-consistent Hartree-Fock-Bogoliubov approximation \cite{Rod} and IBM with \cite{Ramos} and without configuration mixing \cite{McC1,McC2}. It is opportune to remark here about the close similarities between the spectral properties of the $\gamma$-unstable and $\gamma$-rigid collective models \cite{BonZ4,Bud,Rad}, which originate in the common $\gamma$ independence of the collective potential. In the first case it is by choice, whereas in the later it is due to mathematical constraint. On the other hand, some spectral characteristics, for example $\gamma$-band staggering, in some of the heavier Pt isotopes which are historically identified as $\gamma$-unstable can be reproduced only by considering some degree of $\gamma$-rigid \cite{BonZ4,Bug1,Rad} or non-dynamical triaxiality \cite{Yu2,BonZ5,Stag}. $\gamma$-rigid-like triaxial rotations with non-vanishing $K$ but without the compulsory dynamical triaxiality were suggested also for the $N=90$ nuclei \cite{Shaf,Maj}. These arguments show that $\gamma$ rigidity is a useful concept in understanding the shape evolution in the regions of the nuclide chart considered in the present study.

The comparison of theoretical and experimental electromagnetic transitions within the ground state band for these nuclei offers additional information about the evolution of the $\gamma$ rigidity in these two intervals of nuclei. First of all, we must mention the excellent agreement with experiment in the cases of $^{176,178}$Os and $^{182}$Pt. The available experimental data for the Os isotopes are found to be consistent with the evolution of their calculated $\gamma$ rigidities. Indeed, because the $\chi$ value of the $^{180}$Os nucleus, although very little, still deviates from the uniform decrease of $\gamma$ rigidity with neutron number, the corresponding electromagnetic properties do not fall into the present theoretical description. A similar discordance is observed for the already discussed $^{178}$Pt nucleus, while its heavier isotope shows just some signs of redressment. It is worth mentioning that both nuclei are known to exhibit evidences of shape coexistence \cite{McC2,Garcia2}. The $\chi$ value of $^{180}$Pt recommends it as a typical $X(4)$ candidate. However $^{182}$Pt seems a more suitable $X(4)$ representative in the prism of the additional accord between theoretical and experimental transition probabilities \cite{X4}. An interesting situation is also obtained in the $^{184}$Pt nucleus, where the energy spectrum fits prefer high rigidity while the electromagnetic transitions are of the $\gamma$-stable type.

The three sets of nuclei, $N=90$ isotones, Os and Pt isotopes show a regular evolution of the $\gamma$ rigidity. While the $N=90$ nuclei present a $\gamma$-stable valley at $^{152}$Sm and $^{154}$Gd nuclei, the Os and Pt isotopes have monotonous decreasing and respectively increasing of $\gamma$ rigidity as function of neutron number $N$ with some irregular isotopes at the end and respectively start of the considered intervals of nuclei. These findings are also supported by the evolution of deformation in these isotonic and isotopic chains. Indeed, as can be seen from Fig.\ref{b2}, the average deformation calculated within the present model is in a qualitative agreement with the deformation $\beta_{2}$ calculated in the framework of the relativistic mean-field (RMF) theory \cite{Lala}. Due to the scaled nature of the Bohr model results, the average of the $\beta$ deformation is calculated here with Eq.(\ref{bm}), by fixing the position $\beta_{W}$ of the infinite wall to reproduce the RMF $\beta_{2}$ value of the nucleus with the best fit regarding energy spectrum. In this way, the ground state average deformation will depend only on the geometry of the shape phase space by means of parameter $\chi$, just like the spectral properties of the ground and $\beta$ bands.

\section{Conclusions}

The combination of prolate $\gamma$-rigid and $\gamma$-stable rotation-vibration kinetic operators in connection with a flat potential was used to formulate a hybrid critical point model which has as limiting cases the $X(3)$ and $X(5)$ solutions. The relative strength of the rigid and stable components is managed through a so-called rigidity parameter $\chi$. This parameter serves as a weighting measure which bridges the three-dimensional and five-dimensional shape phase spaces of its limiting realizations. The analytical consequences of the shape phase mixing was briefly discussed in general and extensively explained when applied to the critical point solutions. As a result, we obtained a relaxation of the $X(5)$ critical point model through its $\gamma$ rigidity. In this way, we not only improved the agreement with experiment for the well known critical point nuclei but also identified new candidates. Because of the adopted approximations regarding the separation of variables, the energy spectrum of ground and $\beta$ bands of the resulting model depends only on the rigidity $\chi$, while the $\gamma$ band has an additional adjustable energy shift. The numerical applications of the model were directed towards nuclei known to exhibit critical behaviour in the transition from spherical to axially deformed shapes. As a result, many nuclei from $N=90$ isotonic and $Z=76,78$ isotopic chains were found to have non vanishing $\gamma$ rigidity, pointing to a sizable shape phase space mixing. Moreover, specific regularities within its evolution with neutron or proton numbers were evidenced and confirmed by experimental data regarding energy spectra and electromagnetic properties as well as RMF calculations of the ground state deformation.

This result might be used to draw some conclusions about the microscopic structure of these nuclei. Indeed, the $\gamma$ rigidity influences specific mass inertial parameters of the general collective Hamiltonian, which depend on the choice of a particular microscopic nuclear energy density functional or effective interaction when a microscopic theory is mapped into collective variables \cite{Dang, Matsu,Niksic}.

\acknowledgments
The authors acknowledges the financial support received from the Romanian Ministry of Education and Research, through the Project PN-16-42-01-01/2016.

\end{document}